\documentstyle[12pt,aaspp4,flushrt]{article}
\makeatletter

\newlength{\bibhang}
\let\@internalcite\cite
\def\cite{\let\@citeleft(\let\@citeright)%
    \@ifstar{\citeyear}{\citefull}}
\def\citenp{\let\@citeleft\relax\let\@citeright\relax
    \@ifstar{\citeyear}{\citefull}}
\def\citefull{\def\astroncite##1##2{##1~##2}\@internalcite}
\def\citeyear{\def\astroncite##1##2{##2}\@internalcite}

\def\@citex[#1]#2{\if@filesw\immediate\write\@auxout{\string\citation{#2}}\fi
  \def\@citea{}\@cite{\@for\@citeb:=#2\do
    {\@citea\def\@citea{; }\@ifundefined
       {b@\@citeb}{{\bf ?}\@warning
       {Citation `\@citeb' on page \thepage \space undefined}}%
{\csname b@\@citeb\endcsname}}}{#1}}

\def\@cite#1#2{\@citeleft#1\if@tempswa , #2\fi\@citeright}
\def\@biblabel#1{}

\makeatother

\newcommand{\PSbox}[3]{\mbox{\rule{0in}{#3}\includegraphics{#1}\hspace{#2}}}
\newcommand{\FigNum}[1]{\unitlength 1pt \begin{picture}(55,10)(-400,35) 
                        \put(0,0){Figure #1}
                        \end{picture}}
\newcommand{\msun}{$M_\odot$} 
\newcommand{\persec}{\mbox{$\second^{-1}$}}
\newcommand{\percm}{\mbox{$\cm^{-2}$}}
\newcommand{\ppm}{\mbox{$\pm$}}
\newcommand{\cgsflux}{\erg\percm\persec}
\newcommand{\cgslum}{\erg\ \persec}

\newcommand{\approxlt}{\mbox{$\lesssim$}}
\newcommand{\approxgt}{\mbox{$\gtrsim$}}
\def\etal{{et~al.}}

\def\x1608{{4U~1608$-$522}}
\def\cenx4{{Cen~X$-$4}}
\def\nper{{GRO J0422+32}}
\def\nmon{{A0620$-$00}}
\def\nmus{{GS~1124$-$68}}
\def\qzvul{{GS~2000+25}}
\def\v404cyg{{GS~2023+33}}
\def\nsco{{GRO~J1655$-$40}}
\def\ebv{{$E(B-V)$}}
\newcommand{\nh}{\mbox{$N_{\rm H}$}}
\newcommand{\nhtt}{\mbox{$N_{\rm H, 22}$}}
\def\aql{{Aql~X$-$1}}

\newcommand{\ud}[2]{\mbox{$^{+ #1}_{- #2}$}}
\newcommand{\supp}[1]{\mbox{~~$^{#1}$}}
\newcommand{\ee}[1]{\mbox{$10^{#1}$}}
\newcommand{\tee}[1]{\mbox{$\times 10^{#1}$}}

\newcommand{\keV}{\mbox{$\rm\,keV$}}
\newcommand{\MeV}{\mbox{$\rm\,MeV$}}
\newcommand{\cm}{\mbox{$\rm\,cm$}}
\newcommand{\km}{\mbox{$\rm\,km$}}
\newcommand{\second}{\mbox{$\rm\,s$}}

\newcommand{\erg}{\mbox{$\rm\,erg$}}

\newcommand{\rosat}{{\em ROSAT\/}}
\newcommand{\asca}{{\em ASCA\/}}
\newcommand{\rxte}{{\em RXTE\/}}

\newcommand{\beppo}{{\em BeppoSAX\/}}
\newcommand{\kpc}{\mbox{$\rm\,kpc$}}

\newcommand{\kteff}{$kT_{\rm eff}$}
\received{May 27, 1999}
\accepted{September 16, 1999}
\begin{document}
\singlespace
\title{A Method for Distinguishing Between Transiently Accreting 
Neutron Stars and Black Holes, in Quiescence} 

\author{Robert E. Rutledge\altaffilmark{1}, 
Lars Bildsten\altaffilmark{2}, Edward F. Brown\altaffilmark{3}, 
George G. Pavlov\altaffilmark{4}, 
Vyatcheslav  E. Zavlin\altaffilmark{5}}
\altaffiltext{1}{
Space Radiation Laboratory, California Institute of Technology, MS 220-47, Pasadena, CA 91125;
rutledge@srl.caltech.edu}
\altaffiltext{2}{
Institute for Theoretical Physics and Department of Physics, Kohn Hall, University of 
California, Santa Barbara, CA 93111; bildsten@itp.ucsb.edu}
\altaffiltext{3}{
Department of Astronomy and Astrophysics, 
University of Chicago, 
5640 South Ellis Ave, Chicago, IL  60637; 
brown@flash.uchicago.edu}
\altaffiltext{4}{
The Pennsylvania State University, 525 Davey Lab, University Park, PA
16802; pavlov@astro.psu.edu}
\altaffiltext{5}{
Max-Planck-Institut f\"ur Extraterrestrische Physik, D-85740 Garching,
Germany; zavlin@xray.mpe.mpg.de}

\begin{abstract}

Neutron stars and black holes often reside in binaries where the
accretion rate onto the compact object varies by orders of
magnitude. These ``X-ray transients'' are observed both in outburst
(when the high accretion rate makes them X-ray bright) and quiescence
(when the accretion rate is very low, or potentially zero). In a
previous paper, we showed that the quiescent X-ray emission from three
neutron star transients (Aql~X-1, Cen~X-4, and 4U~1608$-$522) were well
represented by thermal emission from the neutron star's hydrogen
atmosphere and that the emitting area was consistent with the whole
surface. Previous black-body spectral fits (which are not accurate
representations of the thermal spectrum) severely underestimated the
true emitting area. In this paper, we fit hydrogen atmosphere models
to the X-ray data for four neutron stars (the three from the previous
paper, plus 4U~2129+47) and six black hole candidates (\nmon, \qzvul,
\nmus, \v404cyg, \nsco, and \nper) with masses $\gtrsim 3\,M_\odot$.
While the neutron stars are similar in their intrinsic X-ray spectra
(that is, similar effective temperatures and emission area radii $\sim
10{\rm\, km}$), the spectra of two black hole candidates are
significantly different, and the spectra of the remaining four are
consistent with a very large parameter space that includes the neutron
stars. The spectral differences between the neutron stars and black
hole candidates favors the interpretation that the quiescent neutron
star emission is predominantly thermal emission from the neutron star
surface.  Higher quality data from {\em Chandra}, {\em XMM}, and
{\em ASTRO-E} will yield a much better contrast.

  There are many transients which do not have clear neutron star
characteristics (such as type~I X-ray bursts or coherent pulsations)
and where the mass of the compact object is not constrained. In these
cases, it is ambiguous as to whether the compact object is a neutron
star or black hole. Our work suggests that an X-ray spectral
comparison in quiescence provides an additional means for
distinguishing between neutron stars and black holes.  The faint X-ray
sources in globular clusters -- thought to be either cataclysmic
variables or quiescent neutron stars -- are a class of objects which
can be investigated in this manner.

\end{abstract}


\section{Introduction} \label{sec:intro}

Many black holes and neutron stars are in binaries where a
steady-state accretion disk (one that supplies matter to the compact
object at the same rate as the mass is donated from the companion) is
thermally unstable \cite{jvp96,king96}.  This thermal instability
results in a limit cycle -- as in dwarf novae -- with matter
accumulating in the outer disk for $\sim$months to decades until a
thermal instability is reached (see \citenp{huang89,mineshige89}), triggering
rapid accretion onto the compact object, and a substantial X-ray
brightening (typically by $\times10^3$ or more, up to $10^{38} \ {\rm
erg \ s^{-1}}$).  Both neutron star (NS) and black hole candidate
(BHC) systems exhibit these X-ray outbursts, separated by periods
($\sim$ months to decades) of relative quiescence (for recent reviews,
see \citenp{tanakalewin95,tanaka96,chen97,campana98b}).

Deep pointed X-ray observations of these transients in quiescence have
found that BHCs are, on average, less luminous in quiescence than
their NS counterparts \cite{barret96,chen97,narayan97a,asai98}.  To
explain the higher quiescent luminosities (\ee{32}--\ee{33}\cgslum) of
the transient NSs, Narayan \etal\ \cite*{narayan97a} suggested that
matter continues to accrete during quiescence onto both objects, and
that neutron stars re-radiate the in-falling matter's kinetic energy,
while black holes swallow most of the accreted mass-energy. In this
picture, the required quiescent accretion rate, $\dot M_q$, onto the
compact object in BHC systems must be substantially greater than in NS
systems.  Current spectral modeling (for an advection dominated flow)
of the few X-ray detected BHC's use $\dot M_q$ of $\sim$ $1/3$ of the
total mass transfer rate in the binary \cite{narayan97b} whereas the
more efficient energy release onto a neutron star requires that $\dot
M_q$ be smaller by 2-3 orders of magnitude.  The cause of this
difference in $\dot M_q$ between the BHC and NS systems is not easily
explained (\citenp{menou99}).

An alternative picture for the NS emission was put forward by Brown,
Bildsten, \& Rutledge \cite*{brown98}, who showed that NSs radiate a
minimum luminosity, even if accretion completely ceases during
quiescence. This minimum luminosity comes from energy deposited in the
inner crust (at a depth of $\sim$300m) during the large accretion
events. The freshly accreted material compresses the inner crust and
triggers nuclear reactions that deposit $\approx {\rm MeV}$ per
accreted baryon there \cite{haensel90}. This heats the NS core on a
\ee{4}-\ee{5} yr timescale, until it reaches a steady-state temperature
$\approx 4\times10^7
(\langle\dot{M}\rangle/10^{-11}\,M_\odot{\rm\,yr^{-1}})^{0.4}\rm\,K$
\cite{bildsten97}, where $\langle\dot{M}\rangle$ is the time-averaged
accretion rate in the binary. A core this hot makes the NS
incandescent, at a luminosity $L_q =
1\MeV\langle\dot{M}\rangle/m_p\approx 6\times 10^{32}
(\langle\dot{M}\rangle/10^{-11}M_\odot\rm\,yr^{-1})\cgslum$, even
after accretion halts \cite{brown98}. The NS is then a thermal emitter
in quiescence, much like a young NS.

For both hypotheses, the energy source for the quiescent luminosities
of the BHCs and NSs have different physical causes, making it
meaningful to search for distinguishing spectral signatures.

\subsection{X-Ray Spectra of Quiescent Neutron Stars}\label{sec:nsback}

The first NS transient detected in quiescence was \cenx4\
\cite{jvp87}.  More recently, quiescent X-ray spectral measurements
have been made of \aql\ \cite{verbunt94} and 4U~2129+47
\cite{garcia99} with the \rosat/PSPC; of EXO~0748$-$676 with Einstein
IPC \cite{garcia99}; and of \cenx4\ and \x1608\ with \asca\
\cite{asai96b}.  The X-ray spectrum of \aql\ (0.4--2.4\keV) was
consistent with a blackbody (BB) spectrum, a bremsstrahlung spectrum,
or a pure power-law spectrum \cite{verbunt94}.  For \x1608, the
spectrum (0.5--10.0\keV) was consistent with a BB ($kT_{\rm BB}\approx
0.2\mbox{--}0.3\keV$), a thermal Raymond-Smith model ($kT =
0.32\ud{0.18}{0.5}\keV$), or a very steep power-law (photon index
$6\ud{1}{2}$).  Similar observations of \cenx4\ with \asca\ found its
X-ray spectrum consistent with these same models, but with an
additional power-law component (photon index $\approx2.0$) above
5.0\keV\ (recent observations with \beppo\ of \aql\ in quiescence also
revealed a power-law tail; \citenp{campana98a}).  The origin of the
observed power-law spectral components in \cenx4\ and \aql\ is not
clear.  While it has been suggested they may be due to magnetospheric
accretion \cite{campana98b}, spectral models of metallic NS
atmospheres \cite{rajagopal96,zavlin96} also predict hard tails.
These warrant further observational investigation.  The observation of
EXO~0748$-$676 shows it to be more luminous (by $\times10-50$) than
the other four NSs. 

In four of these five sources (the exception being EXO~0748$-$676),
BB fits implied an emission area of radius $\approx 1\km$, much
smaller than a NS. This had little physical meaning however, as the
emitted spectrum from a quiescent NS atmosphere with light elements at
the photosphere is far from a blackbody.  For a weakly-magnetic
($B\leq10^{10}$G) pure hydrogen or helium\footnote{The strong surface
gravity will quickly (within $\sim 10\second$) stratify the atmosphere
(\citenp{alcock80,romani87}); for accretion rates $\lesssim 2\times
10^{-13}M_\odot {\rm\,yr^{-1}}$ (corresponding to an accretion
luminosity $\lesssim 2\times 10^{33}\cgslum$), metals will settle out
of the photosphere faster than the accretion flow can supply them
(Bildsten, Salpeter, \& Wasserman \citenp*{bildsten92}).  As a result,
the photosphere should be nearly pure hydrogen. } atmosphere at
effective temperatures \kteff $\lesssim0.5\keV$ the opacity is
dominated by free-free transitions (\citenp{rajagopal96,zavlin96}).
Because of the opacity's strong frequency dependence
($\propto\nu^{-3}$), higher energy photons escape from deeper in the
photosphere, where $T>T_{\rm eff}$
\cite{pavlov78,romani87,zampieri95}.  Spectral fits of the Wien tail
(which is the only part of the spectrum sampled with current
instruments) with a BB curve then overestimate $T_{\rm eff}$ and
underestimate the emitting area, by as much as orders of magnitude
\cite{rajagopal96,zavlin96}\footnote{Application of H atmosphere
models to the isolated neutron stars in SNR~PKS~1209$-$52 and Puppis A
produced a source distance consistent with that measured through other
means (assuming a 10~km NS radius), a lower surface temperature, and
an X-ray measured column density that was consistent with that
measured from the extended SNR (while the column density measured with
an assumed BB spectrum was not consistent with other measurements;
\citenp{zavlin98,zavlin99a}).}.

 Rutledge \etal\ \cite*{rutledge99} showed that fitting the spectra of
quiescent NS transients with these models yielded emitting areas
consistent with a 10~km radius NS.  In Fig.~\ref{fig:nsbbhatm}, we
compare the measured H atmosphere and blackbody spectral parameters
for the quiescent NSs.  The data for 4U~2129+47 is analysed here,
while the other sources were analysed previously
\cite{rutledge99}. The emission area radii are larger from the H
atmosphere spectra by a factor of a few to ten, and are consistent
with the canonical radius of a NS.  There is thus both observational
evidence and theoretical motivation that thermal emission from a pure
hydrogen photosphere contributes to -- and perhaps dominates -- the NS
luminosity at photon energies at 0.1--1 keV. This makes possible the
use of quiescent NS X-ray spectra as an astrophysical tool.

\subsection{Quiescent X-Ray Spectroscopy of NSs and BHCs} 

Distinguishing between a stellar-mass black hole and a neutron star in
an X-ray binary is a non-trivial observational problem. Although there
are X-ray phenomena unique to NSs (for example, type~I X-ray bursts
and coherent pulsations), as yet, no X-ray phenomenon predicted to
occur exclusively in BHs has been observed.  In the absence of any
distinctive NS properties, some X-ray transients are classified as
BHC's if they display X-ray spectral and variability properties
similar to those of other BHCs, such as $\sim$30\% rms
frequency-band-limited variability accompanied by a hard spectrum, or
3-12 Hz quasi-periodic oscillations while the source has high X-ray
intensity, although there are NS systems which display these
properties as well (see \citenp{mvdk95} for a review).  While a
statistical distinction between the quiescent luminosities of BHCs and
NSs has been demonstrated (see refs. in the Introduction), there is
overlap in the observed luminosities; thus, while the different
average quiescent luminosities support the hypothesis that the
(pre-classified) objects belong to distinct classes, quiescent
luminosity cannot be used to distinguish bewteen a NS and a BHC on a
case-by-case basis.  A promising phenomenological distinction between
NSs and BHCs is that, at X-ray luminosities $>$\ee{37} \cgslum, the
20-200 keV luminosity of BHCs is systematically higher than that of
NSs.  However, the physical origin of this difference is not known,
and the weakness of a phenomenological distinction is its
vulnerability to a single counter example \cite{barret96}.  

By far the most solid technique is measuring or constraining the mass
of the compact object via radial velocity measurements of the optical
companion.  If the resulting optical mass function indicates that the
compact object mass exceeds 3\msun, then it is likely to be a black hole,
as the mass-limit of a NS has been calculated to be below 3.2\msun
\cite{rhoades74,chitre76}. The large amount of progress in this method
has given us several very secure black holes \cite{mcclintock98}.

A new spectroscopic distinction between transient NSs and BHs -- based
on the presence of a neutron star's photosphere -- would be a valuable
classification tool. We present the first such comparisons here, where
we report a spectral analysis that uses an accurate emergent spectrum
from a NS to fit the quiescent X-ray spectra of six transient BHCs
with measured mass-functions (\nper, \nmon, \nmus, \nsco, \qzvul, and
\v404cyg) and four transient neutron stars.

 We begin in \S \ref{sec:data} by describing the BHC's and NS's we
have chosen for this comparison, as well as the hydrogen atmosphere
models. Three of the NS are from our previous study \cite{rutledge99},
\aql, \cenx4, \x1608; and we discuss a fourth here, 4U~2129+47. We
show in \S~\ref{sec:comparison}, that although the neutron stars
occupy a narrow range of effective temperatures and emitting area
radii, no such relation is found among the BHCs. This implies that the
H atmosphere spectrum may be used as a tool to distinguish between NSs
and BHCs in quiescence (in the absence of other information).  Section
\ref{sec:energysource} discusses the state of our observational
understanding of the energy source for the quiescent emission from the
NSs. We conclude in \S~\ref{sec:conclude} by summarizing and briefly
discussing the application of this work to X-ray sources in globular
clusters.

\section{Object Selection and Data Analysis}\label{sec:data}

Our purpose in fitting the H atmosphere model -- appropriate only for
NSs -- to the BHC data is to directly compare the measured spectral
parameters of the BHCs to those of NSs.  We selected BHCs for analysis
from among those in Table 1 of Menou \etal\ \cite*{menou99}, which
contains a list of 8 compact binary systems with implied masses
\approxgt 3\msun.  Three of these systems (\nmon, \v404cyg, and \nsco)
were detected in X-rays in quiescence (at luminosities \approxlt
\ee{33} \cgslum), the data for which we analyse; the remaining five
have upper limits in their luminosity.  We analyse the three with
luminosity upper-limits below \ee{33} \cgslum\ (\nper , \qzvul, and
\nmus) to investigate the constraints on emission area radius as a
function of surface temperature.  The remaining two sources with high
luminosity upper-limits $\ge$\ee{33} \cgslum\ (N Oph 1977 and 4U
1543$-$47) are consistent with or greater than the luminosities from
NSs \cite{rutledge99}, and therefore they cannot be excluded as NSs
based on a spectral comparison; we do not investigate them here.

Brief descriptions of the analysed observations are in
Table~\ref{tab:datasets}.  Data were obtained from the public archive
at HEASARC/GSFC \footnote{{\tt http://heasarc.gsfc.nasa.gov/}}.  All
observations analysed are listed in Table~\ref{tab:datasets}; all were
performed with \rosat/PSPC, except for those of \nsco, which were
performed with \asca, and one observation of \nper, which was
performed with \rosat/HRI.

For our spectral fits and derived parameters, we adopted column
densities using the conversion of $\nhtt\equiv N_{\rm
H}/10^{22}\cm^{-2} =0.179~A_V$ \cite{predehl95} and $A_V=3.1~E(B-V)$
\cite{schild77}, except where noted.  We extracted the data for each
source as described in the Appendix, and fit each extracted spectrum
with a spectral model of galactic absorption and a tabulated H
atmosphere model \cite{zavlin96}, using XSPEC \cite{xspec}. 

The H atmosphere model is determined by two spectral parameters: an
effective temperature (\kteff) and the ratio of an emission area
radius ($r_e$; the true radius, which would be measured as the
circumference of the NS, divided by 2$\pi$) to source distance.  In
our results, we quote $r_e$ by assuming a source distance (which can
be uncertain by a factor of 2, and therefore represents a considerable
systematic uncertainty in both the NSs and BHCs).  The H atmosphere
spectrum we used assumes the surface gravity of a 1.4\msun, 10 km
object. It is possible to adopt the surface gravity as an additional
parameter to the H atmosphere model; however, the available data are
of insufficient quality to constrain simultaneously all 3 parameters.

We discuss in the Appendix the column densities adopted for each
individual source.  In quiescence, the S/N of the data is typically
not sufficient to measure the X-ray equivalent column density for the
assumed spectrum.  We thus adopt an historically measured value of
\nh, taken either from: (1) X-ray absorption observed while the object
is X-ray bright; (2) optical reddening, which has been measured
proportional to the equivalent hydrogen column density; or (3) the
neutral hydrogen column density from radio observations
\cite{dickey90} taken from the W3NH tool at HEASARC\footnote{{\tt
http://heasarc.gsfc.nasa.gov/docs/frames/mb\_tools.html}}, which
measures the integrated column density not just to the distance of the
X-ray object, but through the galaxy.  All of these methods, when
applied to estimating the \nh\ during X-ray quiescence, have
systematic errors sufficient to produce $>$100\% uncertainties in the
implied emission area and surface temperature (see discussion in
\citenp{rutledge99}).  In addition, there are observations suggesting
a column density that varies over timescales of months to years,
between outbursts of transients.  An example is a change of
$\Delta\nh=0.5\tee{22}\cm^{-2}$ in 4U~1608$-$522 \cite{penninx89} over
several months.  This uncertainty in \nh\ can only be overcome by high
S/N data in X-ray quiescence, which can permit a direct measurement of
\nh\ during the observation.  Interpretation of the results of the
present analysis must bear this systematic uncertainty in mind.

\section{Comparison between BHCs and NSs in Quiescence}
\label{sec:comparison}

In Fig.~\ref{fig:nsbhcs}, we compare the measured spectral parameters
\kteff\ and $r_e$ of quiescent NSs and BHCs.  Error bars are 90\%
confidence, as are upper-limits.  For the four NSs in this analysis,
the \kteff\ and $r_e$ were constrained, with best-fit \kteff\ in the
range 0.08--0.20~keV, and $r_e$ in the range 8--12~km. The NSs exhibit
a significant range in temperature, possibly due to different core
temperatures (related to $\langle\dot{M}\rangle$).  The data are of
sufficient quality to constrain $r_e$ to within a factor of 2 for the
NSs (not accounting for the uncertainties in source distance), and the
resulting $r_e$ are consistent with objects of 10~km radius.  The two
connected points for \aql\ are for two different distances (2.0 and
4.0 kpc).  The two connected points for 4U~2129+47 are discussed in
the analysis section~\ref{sec:2129}.  We note that recent observations
\cite{callanan99} have resolved the optical counterpart of \aql\ into
two objects, only one (at most) will be associated with the X-ray
system; most likely, this observation implies a distance to \aql\
greater than previous estimates, although a new distance estimate has
not yet been produced.

The spectra of five of the six BHCs were of insufficient statistics to
simultaneously constrain both $r_e$ and \kteff\ of the H atmosphere
model.  The exception (\v404cyg) was significantly harder than the NS
spectra, which constrained the \kteff\ to be above those observed from
the NSs.  The BHCs \nsco, \qzvul, \nper\ and \nmus\ have spectral
parameters which are, within errors, consistent with those of the NSs.
\nmon\ and \nper, if we presumed these to be NSs of the same surface
area and gravity, must be cooler (\approxlt 0.05~keV) than the average
observed transient NSs.  The spectrum of \nmon\ is discrepant with those
of the NSs, implying substantially smaller emission area
($r_e\sim0.2\mbox{--}2\rm\,km$) for the range of \kteff\ observed from
the NSs.

The addition of a power-law spectral component -- such as the hard
power-law tail as observed from Cen X-4 in {\it ASCA} data, and
detected in Aql~X-1 in BeppoSAX data -- introduces enough uncertainty
in the H atmosphere spectral parameters of BHCs that all sources would
be consistent with those we observe from the NSs.  In noting this
systematic uncertainty, we point out that additional spectral
components are not demanded by the data we have used, although the
data is largely of low-bandwidth (0.4-2.4 keV), and that wider
bandwith data (such as from {\it ASCA}, {\it Chandra}, or {\it XMM})
may alter this.  In practice, we have neglected this systematic
uncertainty to investigate the question: if these BHC sources were
discovered in quiescence, and their H atmosphere spectral parameters
were compared with those from known NSs, would we conclude they are
similar or dissimilar to the NSs?  For GS~2023+33 and A0620$-$00, we
find that they are dissimilar; while for GRO J1655$-$40, GS1124$-$683,
GS2000+25, and GRO J0422+32, we find they are consistent (albeit with a
wide range of values as well).  Higher signal-to-noise data, taken
with an instrument of wider bandwidth, would permit us greater
certainty regarding the possible contributions of spectral parameters
for which we have not here accounted.

Moreover, no BHC has well-constrained spectral parameters which would
place it exclusively within the parameter space occupied by the NSs,
as depicted on Fig.~\ref{fig:nsbhcs}.  This is a phenomenologically
defined region, which would have to be expanded with the discovery of
NSs in quiescence outside of this box.  Thus, we find no evidence that
any of the BHCs have been misclassified, and should be re-classified
as NSs.

\section{The Energy Source for the Neutron Star Quiescent Luminosity}
\label{sec:energysource} 

There are presently no observational results which exclude that part
of the quiescent luminosity of these NSs is due to accretion.  Brown
et al. (1998) noted a few observational tests which can be applied to
determine whether accretion is active.  First, as accretion will
increase source luminosity (at this low luminosity level), the spectra
should be drawn when sources are at their lowest observed luminosity
in the X-ray passband.  Second, since the H atmosphere thermal flux is
expected to be variable only on timescales of longer than $\sim$months
to $10^4$ yr, variability on timescales less than this (\approxlt
days) likely indicates active accretion. Second, active accretion onto
the NS surface will produce metal absorption lines in the spectrum (O
and Fe, below 1 keV), which can be observed with {\it Chandra} and
XMM, but which cannot be observed with {\it ASCA} or {\it ROSAT}. The
presence of photospheric absorption metal lines in the spectrum --
aside from being observationally important -- will indicate active
accretion onto the NS surface. These indicators should be used to
insure that the observed NS emission is not due to active accretion
(in addition to being observationally consistent with the theoretical
H atmosphere spectrum).  Since present instrumentation is not capable
of detecting the lines, we can only apply the variability and observed
luminosity criteria.  

If accretion is occuring during quiescence, this will increase the
\kteff\ of an emergent H atmosphere spectrum, and produce metal
absorption lines; accretion will not affect $r_e$, unless emission
originates from a different surface than the NS (such as in an
accretion disk).  

Some observational evidence suggests that accretion is indeed
occurring onto the NS surface during quiescence; long-term
(months-years) variability in the observed flux has been reported (for
4U~2129+47, see App.~\ref{sec:2129var} and \citenp{garcia99}; for
\cenx4, \citenp{jvp87}).  While this variability can be explained by a
variable absorption column depth, active accretion during quiescence
is also a possibility.  However, recent observations of \aql\ at the
end of an outburst showed an abrupt fading into quiescence
\cite{campana98a} associated with a sudden spectral hardening
\cite{zhang98a}. This was followed by a period of $\sim$15 days, over
which the source was observed (three times) with a constant flux level
\cite{campana98a}.  This behavior was interpreted as the onset of the
``propellor effect'' \cite{ill75,stella86} in this object, which would
inhibit -- perhaps completely -- accretion from the disk onto the NS.
The energy source for the long-term nearly constant flux is then a
puzzle.  Unlike the accretion-powered models, the work of Brown et
al. (1998) makes a specific prediction for this constant flux level,
relating it to the long term time-averaged accretion rate.

So as to minimize any contributions from accretion, we only analyse
observations made during periods of the lowest observed flux.  We
calculate the bolometric luminosity from the H atmosphere fits.  The
fits are given in terms of an (unredshifted) effective temperature
(\kteff) and emitting area radius ($r_e$).  The observed bolometric
flux is $F_{\rm bol} = (1+z)^{-2} (r_e/d_0)^2\sigma T_{\rm eff}^4$,
where $d_0$ is the source distance, and the observed bolometric
luminosity is $L_{\rm bol}=4\pi d_0^2 F_{\rm bol}$.  For a NS of
$1.4\,M_\odot$ and $10\rm\,km$ radius, the surface redshift is
$1+z=(1-2GM/Rc^2)^{-1/2}=1.31$.  In Table~\ref{tab:lbol} we give the
unabsorbed, observed X-ray and bolometric luminosities assuming a H
atmosphere spectrum and the surface redshift of a $1.4\,M_\odot$
$10\rm\,km$ radius NS ($1+z=1.31$), measured as described in the
analysis of the present work and previously \cite{rutledge99}.  The
luminosity quoted for \x1608\ in the previous work was not calculated
for the distance stated there; the correct value is in this table.
While there is considerable systematic uncertainty in these values,
due to uncertain distances and \nh, the luminosities cover an order of
magnitude, at about \ee{32-33} \cgslum.

Using these new bolometric quiescent luminosities for Aql~X-1,
Cen~X-4, and 4U~1608$-$522, we have remade a plot (Fig.~\ref{fig:Lq})
of $L_q/L_o$ as a function of $t_r/t_o$ \cite{brown98}.  Here $L_q$
and $L_o$ are the observed quiescent and average outburst
luminosities, and $t_r$ and $t_o$ are the recurrence interval and
outburst duration.  We show this relation for the NSs ({\em open
circles\/}) Aql X-1, Cen X-4, 4U~1608$-$522, and EXO~0748$-$676 and
the BHCs ({\em filled circles}) H~1705$-$250, 4U~1543$-$47, Tra~X-1,
V~404~Cyg (GS~2023+33), GS~2000+25, and A~0620$-$00.  We denote with
an arrow those BHCs for which only an upper limit on $L_q$ is known.
Because the recurrence time for the NS 4U~2129+47 is unknown, we do
not show it here.  The expected incandescent luminosity is plotted for
two different amounts of heat per accreted nucleon stored in the core
during an outburst: 1\MeV ({\em solid line}) and 0.1\MeV ({\em dotted
line}).  The use of bolometric quiescent luminosities moved Aql~X-1,
Cen~X-4, and 4U~1608$-$522 to higher $L_q/L_o$ (upwards on this
diagram).  With the exception of these three objects and the Rapid
Burster ($L_q$ is from \citenp{asai96a}), the data from this plot is
taken from \cite{chen97}.  For Aql~X-1 and the Rapid Burster, $L_o$
and $t_o$ are accurately known (\rxte/All-Sky Monitor public data);
for the remaining sources $L_o$ and $t_o$ are estimated from the peak
luminosities and the rise and decay timescales.

Four of the five NSs are within the band where the quiescent
luminosity is that expected when the emitted heat is between 0.1-1.0
MeV per accreted baryon.  The fifth NS (EXO~0748$-$676), has a higher
quiescent luminosity (by a factor of 10), which we interpret as being
due to continued accretion, an interpretation which is reinforced by
the observation of spectral variability during the quiescent
observations with {\it ASCA} \cite{corbet94,thomas97}, on timescales of
$\sim$1000 sec and longer. (Garcia and Callanan \citenp*{garcia99}
measured $L_x$=1\tee{34} \cgslum\ from Einstein/IPC observations of
this source.)  The BHCs on this figure are more spread out across the
parameter-space, qualitatively indicating a statistical difference --
although not one which is particular for each object -- between the
two classes of objects.  This suggests the NS quiescent luminosity is
more strongly related to the accreted energy than the BH quiescent
luminosities.

\section{Conclusions}\label{sec:conclude}

We have fit the quiescent X-ray spectra of four transient NSs and six
transient BHCs (with measured mass functions) with a pure H atmosphere
spectrum.  We compared the emitting area radius and effective
temperature of the ten sources and found that the NSs are clustered in
($r_e$, \kteff) parameter space.  Two of the BHCs (\nmon, \v404cyg)
are inconsistent with the NS spectra; the upper-limits of three more
(\nmus, \qzvul, \nper), and the ($r_e$, \kteff) locus of the sixth
(\nsco) overlap the NS parameter space.  The upper-limits of the
parameter space of \qzvul\ and \nper\ are marginally consistent with
the observed ($r_e$, \kteff) of the NSs, indicating that these BHCs,
if interpreted as NSs, would have to be cooler than the average NS
observed in quiescence.

We found that the X-ray spectra excludes \nmon\ and \v404cyg\ from
being NSs of the type we used for comparison.  These differences
between the quiescent X-ray spectra of BHCs and NSs cannot be directly
attributed to an observational selection effect.  The NSs were
identified by the type~I X-ray bursts, the BHCs were first identified
by their X-ray variability behavior and later by the high (\approxgt
3\msun) implied mass of the compact object.  The X-ray spectra of
\nmus, \qzvul, \nsco, and \nper\ are consistent with those of the NSs,
but their parameter space is large, and better data are needed to
determine whether they are spectrally similar or not.

Our method can be applied to objects which have low X-ray luminosities
(\approxlt $10^{34}$ \cgslum), to identify them as neutron stars,
which have an atmosphere from which the theoretical emission
originates, or as some object which is not a NS, as we have done here.
In addition to the transient field objects, this method can be readily
applied to the low-luminosity X-ray sources observed in globular
clusters \cite{hertz83} which are thought to be cataclysmic variables
\cite{cool95,grindlay95}, but may also be transient neutron stars in
quiescence \cite{verbunt84}.  X-ray spectroscopic determination can
identify these objects as NSs radiating thermal emission from the
atmosphere, or imply a different origin for the emission. As discussed
above and elsewhere \cite{brown98}, the quiescent luminosities of
these sources are set by the time average accretion rate. Thus, the
low luminosity ($10^{31}$ \cgslum) X-ray sources in globular clusters,
if they were transient neutron stars in quiescence, would have
$\langle \dot{M} \rangle \approx 2\tee{-13}$ \msun $\ {\rm yr}^{-1}$;
comparing this to Aql~X-1, with $\langle \dot{M} \rangle \approx
1\tee{-10}$ \msun $\ {\rm yr^{-1}}$ (estimated from the RXTE/ASM
lightcurve history), and a mean outburst interval of $\approx$ 200
days, the low-luminosity X-ray sources would have Aql--like outbursts
with recurrence times of $\sim$250 yr, assuming the quiescent
luminosity is in steady-state with accretion.  The estimated number of
such sources down to this luminosity level is highly uncertain, about
1-10 per globular cluster \cite{verbunt95}.  For $\sim$200 GCs in the
galaxy, one expects $\sim$1-10 such transients per year. Even assuming
only 1/4 of all GCs are within distance to detect an Aql~X-1-like
transient with the RXTE/ASM (for a peak RXTE/ASM detection countrate
of 10 c/s; Aql~X-1 has an ASM peak countrate of $\sim$30 c/s), this
produces an expected transient discovery rate of 0.3-3 per year, which
is consistent with or greater than the observed discovery rate with
RXTE/ASM (of no sources over a four-year period).  Stronger
constraints could be made with more sensitive all-sky monitoring
observations over longer time-baselines, or with a more tightly
constrained luminosity function of the low-luminosity sources in
globular clusters.

Higher quality X-ray data from the coming X-ray spectroscopy missions
({\it Chandra/ACIS, XMM} and {\it ASTRO-E}) will permit this analysis
to be performed with greater accuracy, in particular by permitting the
simultaneous measurement of the X-ray column density---a dominant
systematic uncertainty.  These will also provide the means to account
for possible contributions due to a hard-power law component in the
BHCs. For example, a 15~ksec observation with {\it Chandra} of a
source with photon power-law slope of 2, and luminosity of 2\tee{31}
\cgslum , \nhtt=0.2 at 1kpc would produce a spectrum which can be
excluded as a H atmosphere (or blackbody) with an (unconstrained)
column depth, with probability=6\tee{-6}.  Finally, the high spectral
resolution and countrates of these instruments will permit a search
for short timescale variability and photospheric metal absorption
lines, which would indicate ongoing accretion during X-ray quiescence
\cite{brown98}.

\acknowledgements

This research was supported by NASA via grant NAGW-4517 and through a
Hellman Family Faculty Fund Award (UC-Berkeley) to LB, who is also a
Cottrell Scholar of the Research Corporation.  EFB is supported by
NASA GSRP Graduate Fellowship under grant NGT5-50052. GGP acknowledges
support from NASA grants NAG5-6907 and NAG5-7017.  This research was
supported in part by the National Science Foundation under Grant No.
PHY94-07194.  We acknowledge use of data obtained through the High
Energy Astrophysics Science Archive Research Center Online Service,
provided by the NASA/Goddard Space Flight Center.  We gratefully
acknowledge useful conversations with D. Fox and A. Prestwich
regarding \rosat/HRI response issues.  We gratefully acknowledge
helpful comments from J. McClintock and from the referee, J. Grindlay.

\appendix

\section{Sources}\label{sec:sources}

The assumed distances (used to calculate luminosity) and adopted \nh\
for the spectral fits performed here are listed in
Table~\ref{tab:objects}; values for the NSs we analysed previously
(\aql, \cenx4, \x1608) are in the previous reference
\cite{rutledge99}.  The results of the spectral fits are presented in
Table ~\ref{tab:results}, which contains: (1) the dataset number
(cf. Table~\ref{tab:datasets}); (2) the \nh; (3) the best-fit spectral
parameters for the H atmosphere model, including the {\it
un}-redshifted effective NS surface temperature (\kteff) and apparent
emission area radius $r_e$(km), as well as the reduced $\chi^2_\nu$
for that model.  The H atmosphere model -- appropriate only for NSs --
is applied here to the BHCs, to produce spectral parameters which can
then be directly compared with those observed from NSs.

Here, we describe our assumptions about individual objects, and
provide specifics of the analyses of each observation.

\subsection{\nper}

The color excess, measured from an IUE spectrum of this object, was
found to be \ebv=0.40\ppm0.07 (\nhtt$\sim$0.23 \ppm0.04; \citenp{shrader92}).

First, we analyse observation 1, a ROSAT/PSPC observation which had
not previously been analysed (this observation is included for
completeness, as this source was fainter during observation 2).  We
extracted the data from a 53\arcsec\ circle about the source, and
background from a 530\arcsec\ annulus about the source, excluding
another object in the FOV, and 3 low-surface brightness areas
clustered in the SE side of the annulus (possibly background
fluctuations).  Due to the long exposure, this spectrum constrained
all three parameters (\nh, \kteff, and $r_e$).  The resulting best-fit
was $r_e$=3.0\ud{2.4}{0.7} km, \kteff=0.22\ud{0.03}{0.05} keV, and
\nhtt$<0.12$ (90\%), for an implied luminosity of 7.8\tee{32} (d/2.6
kpc)$^2$ \cgslum\ (0.5-2.0 keV).  In a fit holding the column density
fixed at \nhtt=0.22, the best fit model was not acceptable at the
p=1.8\% level.  The parameter space for a $\Delta \chi^2$=2.72 would
constrain the $r_e$ to be similar to those found from the NSs
(11\ud{4.7}{2.5} km).

We analysed the HRI observation 2.  The observation was previously
analysed, producing a luminosity upper-limit of log(L)$<$31.6 (d=2.6
\kpc, $\alpha$=2.1, and \nhtt=0.2; \citenp{garcia97}).  We extracted
the spectrum from an 8\arcsec\ radius about the source position --
finding 7 photons, consistent with the expected and measured
background count-rate. We extracted background counts from an annulus
of 50\arcsec\ and 10\arcsec\ outer and inner radii, respectively.  We
used the Dec 1 1990 HRI spectral response from GSFC/HEASARC
calibration database.  We find a slightly lower, but consistent
$3\sigma$ upper-limit to the unabsorbed source luminosity for the same
assumed spectrum ($\log(L)<31.4$) as found previously.  We assumed a
series of temperatures and found upper-limits to $r_e$ for the assumed
spectrum.  For a comparable temperature as derived from the PSPC
analysis (above), the implied $r_e<0.8$ km -- substantially below that
found from the PSPC analysis, which indicates intensity variability at
this low luminosity.  As observation 2 has lower luminosity than
observation 1, we use observation 2 in our interpretations (as thermal
emission from the NS surface should not vary by \approxgt few\% over
the timescales here, and it is only the lowest observed luminosity
from each object which may be due in its greatest part to the thermal
surface emission).

\subsection{\nmon}
The optical color excess of the companion star has been measured as
\ebv=0.39\ppm0.02. \cite{wu76,oke77}, corresponding to \nhtt=0.22 --
somewhat below the \nhtt=0.41 \cite{dickey90} measured in this
direction (which is the integrated value along this line of sight
through the entire galaxy), consistent with a nearby object. We adopt
the \nhtt=0.22 value for our spectral fits.

We analysed data taken by \rosat/PSPC (observation 3); this data was
previously analysed \cite{mcclintock95}.  An X-ray source was found in
the extracted data at the source position.  Counts from this source
were extracted from a circle of radius 53\arcsec.  Background was
taken from an annulus about the source, 4\arcmin\ outer radius, and
60\arcsec\ inner radius; excluded from this annulus were circular
areas (each only half-within the annulus, straddling the outer radius)
about two unrelated sources, each with radius 53\arcsec.  The source
region contained a total of 116 counts. The background region contained
938 counts.  We used data in the energy range 0.4-2.1 keV, rebinned
into three energy bins 0.4-0.8, 0.8-1.2, and 1.2-2.1 keV.  Using the
same spectral model and assumptions as the previous work
\cite{mcclintock95}, we reproduce the best-fit blackbody temperature
and source luminosity.

The data are of insufficient S/N to constrain the temperature and area
independently.  We held the temperature fixed at values
\kteff$=[0.04,0.54]$ keV, and extracted the best fit emission area
radii ($r_e$), which were found to range from 14 \ppm 2 km (at
\kteff$=0.04$ keV) to 0.07 km (at \kteff$=0.15$ keV). The model
$\chi^2_\nu$ reaches a probability of 2\% that the observed spectrum
is produced by the model in a single random trial ($\chi^2_\nu=4.0$,
for 2 dof) at \kteff$=0.265$ keV, which sets our upper-limit on the
effective temperature for this source.

\subsection{\nmus}

The X-ray measured equivalent column density was found to have
evolved, decreasing by \nhtt$\sim$0.2 during the outburst, and
settling to a value of \nhtt=0.16 \cite{ebisawa94}, consistent with
the optical reddening \ebv=0.20\ppm 0.05 \cite{gonzalez91} measured
from a UV absorption line, and is below the measured hydrogen column
density in the direction of this source \nhtt=0.25 \cite{dickey90},
consistent with an object a short distance away relative to the size
of the galaxy. We adopt \nhtt=0.16 for our spectral fits.

The PSPC observation (number 4) has been previously analysed
\cite{greiner94,narayan97a}.  For the source, we extracted counts from
a 75\arcsec\ radius circle about the source position, as the source
position is about 15\arcmin\ off-axis.  For background, we used three
circular areas on the detector, one centered at the source position,
the other two centered at distances from the FOV center equal to that
of the source position.  From these three circular areas, we excluded
the 75\arcsec\ region about the source position, and one other
apparent (that is, faint) source. There were 66 counts in the source
region, and 2104 counts in the background spectrum.  We generated an
ancillary response file for this off-axis spectrum using the FTOOL
pcarf V2.1.0.  In the spectral analyses, we find a 3$\sigma$
upper-limit to the (unabsorbed) source flux of $<$3.5\tee{-14}
\cgsflux (0.3-2.4 keV), assuming \nhtt=0.22, and a power-law photon
spectral slope of $\alpha=2.5$, consistent with the previously found
value.

There is no detectable quiescent source at the position of the optical
source, consistent with previous results \cite{greiner94,narayan97a}.
Across the investigated range of NS surface temperatures, the
upper-limits of the implied radii range from 82-0.13 km.

\subsection{\nsco}

This \asca\ observation has been analysed previously \cite{ueda98}, in
which the flux was measured to be 2\tee{-13} \cgsflux\ (2-10 keV).  The
interstellar column density was measured at \nhtt=0.74\ppm0.03 in the X-ray
high (bright) state, and $<$0.08 and $<$0.14 (90\%)in the low (faint)
state.  The optical reddening is \ebv=1.3\ppm0.1
(cf. \citenp{orosz97,horne96}) is consistent with a value of
\nhtt=0.74, which we adopt and hold fixed.

Using data taken with the GIS in PH mode, we extracted background
counts from two circles, each 5\arcmin\ in radius and centered
5\arcmin\ from the image center -- the same distance as the position
of the object.  The source X-ray spectra were extracted from a
5\arcmin\ circle about the optical position.  We excluded energy bins
below 1.0 keV and above 5.0 keV from the fit.

For the SIS, data were taken from 2.5\arcmin\ radius circles about the
source.  Background was taken from a square approximately 6\arcmin\ on
a side, excluding the 2.5\arcmin\ source region.  We used energy range
of 1.0-10.0 keV for the GIS data, and 0.5-10.0 keV for the SIS data. 

Based on this data, no single best fit is found for $0.04 {\rm
keV}<$\kteff$<0.54 {\rm keV}$.  As the assumed temperature (held fixed) is
increased from \kteff$=$0.04 to 0.54 keV, $r_e$ decreases from 110\ppm32
km to 0.19\ppm0.02 km.

\subsection{\qzvul}

The X-ray equivalent \nh\ for this object has been measured as
\nhtt=1.14 \cite{tsunemi89}, and optical reddening measured as
\ebv=1.5 \cite{chev90} (corresponding to \nhtt=0.825), which are
slightly discrepant from one another, though within the systematic
uncertainties between these techniques.  The measured hydrogen column
density in the direction of this object is \nhtt=0.66 \cite{dickey90},
although within a degree, values range between \nhtt=[0.51,0.81], and
at the most proximate point, \nhtt=0.67.  The higher measured optical
reddening of the stellar companion and absorption of X-rays from the
compact object, relative to the direct \nh\ measurement, may indicate
significant absorption in the local environment of the binary, on the
order of \nhtt=0.5.  We use in turn \nhtt=0.66 and \nhtt=1.1.

Two of the presently analysed observations (\#6 \& 7) have been
analysed previously \cite{verbunt94}.  In the \rosat\ image for any of
the three observations (\#6, 7, and 8) there is no detected object at
the source position.  We combine the data, which are of three widely
varying epochs, to produce average upper limits to spectral
parameters.

In each of the three observations, we extract source counts from a
circular region centered on the object of 60\arcsec\ radius.
Background counts were taken from annuli about the source, with an inner
radius of 63.75\arcsec\ and an outer radius of 375\arcsec.  We also
excluded two circular regions (60\arcsec\ radius) which overlapped the
annuli, which appear to contain sources.  There were a total of 78
counts in the source region, and 2505 counts in the background region.
The resultant spectrum produces flux upper-limits consistent with (but
below, by about 25\%) the higher upper-limits found previously
\cite{verbunt94}.

We do not detect a source above the background countrate; we produce
90\% upper limits to $r_e$, which were found to range from 0.13-28 km,
depending on the assumed surface temperature, in the range of
0.04-0.54 keV.

\subsection{\v404cyg}

The optical reddening has been measured at $A_V\sim 3.0$
\cite{charles89}, and a comparable \ebv=1.03 \cite{wagner91}, which
imply \nhtt=0.54-0.59. This is below the measured \nh=0.81 in this
direction \cite{dickey90} consistent with an object nearby relative to
the size of the galaxy.   We adopt a value of \nhtt=0.54. 

This object was observed twice in quiescence: in 1994 with \asca\ and in
1992 with \rosat/PSPC (see \citenp{verbunt94}).  It was also observed
and analysed in the \rosat\ All-Sky Survey \cite{verbunt94}, on Nov 5,
with an average countrate of 0.028\ppm 0.009 c/s,in which it was found
with a luminosity of about 1.5\tee{33} (d/3.5 \kpc)$^2$ \cgslum\
(0.4-2.4 keV).

From the ROSAT image, data were selected centered on the X-ray source,
in a circle 60\arcsec\ in radius. Background was taken from an annulus
with an inner-radius of 63.8\arcsec\ and an outer radius of
375\arcsec .  The source region had 425 counts, and the background
region had a total of 1365 counts. 

At effective temperatures below, \kteff$=0.258$, the model becomes
untenable ($\chi^2_\nu$=4.0, 2 dof, corresponding to 98\%
probability), which sets the lower-limit of the acceptable effective
temperature parameter space for this data.  The best fit area was not
well constrained from below (as the temperature increases to values
above those for which this model is valid), but the 90\% upper limit
on the radius is 1.9 km (d/3.5 \kpc).  We find the (0.4-2.4 keV)
luminosity of the best fit model to be 6\tee{32} $(d/3.5 {\rm
kpc})^2$\cgslum .

These results are consistent with those of an earlier analysis of
these same data \cite{wagner94}, although we note that intensity
variability was found on $\sim$days timescales which, {\it a priori}
cannot be attributed to the NS surface flux from the hot (thermal)
core.  These results are also consistent with results from \asca\
analysis of quiescent data taken at a different time
\cite{narayan97b}.

\subsection{4U 2129+47}
\label{sec:2129}
We inadvertently did not include this NS transient in our previous
study, so we analyse it here.  This spectrum has been analysed
previously with a black-body model \cite{garcia99}, and the derived
parameters were consistent with those we find here, under similar
distance/column density assumptions. 

The distance to this object is controversial with values between 1 and
6 kpc (see \citenp{cowley90} for arguments on both sides).  Associated
with this distance uncertainty is an optical reddening uncertainty; if
the optical counterpart is more distant, it must be more luminous, of
earlier type, and bluer, in which case the reddening is \ebv=0.3
(\nhtt=0.17; \citenp{cowley90}); if less distant, the companion is
redder, and \ebv=0.5 (\nhtt=0.28; \citenp{thorstensen79,chev89b}).
The total \nh\ measured in the direction of this object is \nhtt=0.38
\cite{dickey90}, while the best fit column density obtained by Garcia
and Callanan (with this same data; \citenp*{garcia99}) was \nhtt=21.5
(no quoted error).

In interpreting the data, we alternately adopt a distance of 1.5 kpc
and \ebv=0.5 (\nhtt=0.28), and 6.0 kpc and \ebv=0.3 (\nhtt=0.17).

For the March 1994 PSPC observation, the source counts were extracted
from a circular region 45\arcsec\ in radius, with a total 279 counts.
Background was extracted from an annulus about the source, 278\arcsec\
and 45\arcsec\ in outer and inner radius, respectively, for a total of
1959 counts.  We fit to data in the 0.4-2.4 keV energy range.

For an assumed \nhtt=0.28 (d=1.5 kpc), the best fit parameters were
\kteff$=$0.08\ud{0.02}{0.015} keV, and $r_e$=7.1\ud{6.0}{3.5} km.  For
\nhtt=0.17 (d=6.0 keV), the best fit parameters are
\kteff$=$0.105\ud{0.025}{0.020}, and $r_e$=12.0\ud{8.6}{4.9}.  The
unabsorbed luminosities for the assumed (\nh, d) combinations are
6.5\tee{31} and 5.6\tee{32} \cgslum\ (0.4-2.4 keV), respectively.

\subsubsection{An Observed Change in the Flux from 4U 2129+47}
\label{sec:2129var}
To compare the observed flux of 4U~2129+47 analysed here, with one
measured previously with the \rosat/HRI during a 7 ksec of
observations (Nov-Dec 1992; \citenp{garcia97}), we imposed a \nhtt=0.5
and a photon power-law slope of 2.0, and measured the (absorbed) flux
from the March 1994 observation, which was (6.7\ppm0.5)\tee{-14}
\cgsflux (0.3-2.4 keV) -- a factor of 3.4\ppm0.6 below that measured
during the earlier observation by Garcia (we independently analysed
the earlier HRI observation, and confirm Garcia's flux
measurement). We cannot simultaneously extract $r_e$ and \kteff\ from
the HRI observation, as only one spectral bin is available, and so we
cannot compare the photon energy spectra. However, we jointly fit both
observed spectra with the same H atmosphere and \nh , and found that
there is a 2.7\% chance that the best-fit spectrum produced both
observed spectra.  Assuming that $r_e$ and \nh\ are the same between
the two observations, the effective temperature would have to have
dropped from 0.091\ppm0.006 keV (90\%) to 0.081\ppm0.002 keV to
explain the different spectra.  The thermal timescale at the depth
where the crust heating occurs is about a year \cite{brown98}; the
decrease in effective temperature could be a signature of a cooling
crust.  Another possible explanation is that the \nh\ was lower during
the 1992 observation than the 1994 observation; for example if
\nhtt=0.10 during the HRI observation and \nhtt=0.28 during the 1994
observation, but with the same underlying H atmosphere spectrum, this
would account for the observed difference in flux. In addition, if
accretion is active during quiescence, the different X-ray spectra may
reflect a decrease in the accretion rate by a factor
(0.081/0.0906)$^4$=0.64, assuming a thermal spectrum.

With the quality of spectra we have here, we cannot definitely state
the cause of the discrepancy between the 1992 and 1994 observations,
but it is consistent with either a difference in the intervening \nh,
a difference in the effective temperature, time-variable accretion, or
a combination of the three.

\clearpage

\begin{figure}[htb]
\caption{ \label{fig:nsbbhatm} Comparison between the spectral
parameters $r_e$ and \kteff, derived from spectral fits of the
quiescent X-ray emission from \aql, \cenx4, \x1608 and 4U~2129+47.
The ({\em open points}) are for the H atmosphere spectrum and the
({\em solid points}) are from a black-body spectrum.  The two points
connected for \aql\ correspond to the upper- and lower- distance
limits for that source.  The two connected points for 4U~2129+47 are
for the two different distance/\nh\ estimates (see text).  As expected
from comparisons between the theoretical spectra, the H atmosphere
fits produce higher values of $r_e$. }
\end{figure}


\begin{figure}[htb]
\caption{ \label{fig:nsbhcs} The best fit emission area radii for NSs
and BHCs (\nmon, \v404cyg, \nsco) using a hydrogen atmosphere
spectrum, with the surface gravity of a 1.4\msun, 10km object.  The
dotted-line box encloses the (phenomenological) parameter space
occupied by the NSs.  The 90\% upper-limits for \qzvul, \nmus\ and
\nper, are not included (see Table~\ref{tab:results}), but are lines
which lie between \nsco and \nmon across \kteff=0.04--0.054 keV,
passing through the NS box.  The best-fit parameters for the NSs
cluster in the \kteff$\sim$0.08-0.2 keV and $r_e\sim$8-12 km.  The
best-fit parameters for BHCs \nsco, \nmus, \qzvul, and \nper\ are
consistent with having similar spectra, while \v404cyg\ and \nmon\ are
inconsistent.  }
\end{figure}

\begin{figure}[htb]
\caption{ \label{fig:Lq} The ratio of outburst luminosity $L_q$ to
quiescent luminosity $L_o$ as a function of the ratio of recurrence
interval $t_r$ to outburst duration $t_o$.  The lines are for different
amounts of heat, $0.1\MeV$ ({\em dashed line\/}) and $1.0\MeV$ ({\em
solid line\/}),  per accreted nucleon deposited at depths where the
thermal time is longer than the outburst recurrence time.  Also plotted
are the observed ratios for several NSs ({\em filled squares\/}) and BHs
({\em open squares}).  For most of the BHs, only an upper limit ({\em
arrow\/}) to $L_q$ is known.  Data is from \protect\cite{chen97},
with the exception of $L_q$ for the Rapid Burster
\protect\cite{asai96b}.  For Aql~X-1 and the Rapid Burster, $L_o$ and
$t_o$ are accurately known (\rxte/All-Sky Monitor public data); for the
remaining sources $L_o$ and $t_o$ are estimated from the peak
luminosities and the rise and decay timescales.  For Aql~X-1, Cen~X-4,
and 4U~1608$-$522, we have recalculated the bolometric luminosities using
the H atmosphere models \protect\cite{rutledge99}.}
\end{figure}

\clearpage
\pagestyle{empty}
\begin{figure}[htb]
\PSbox{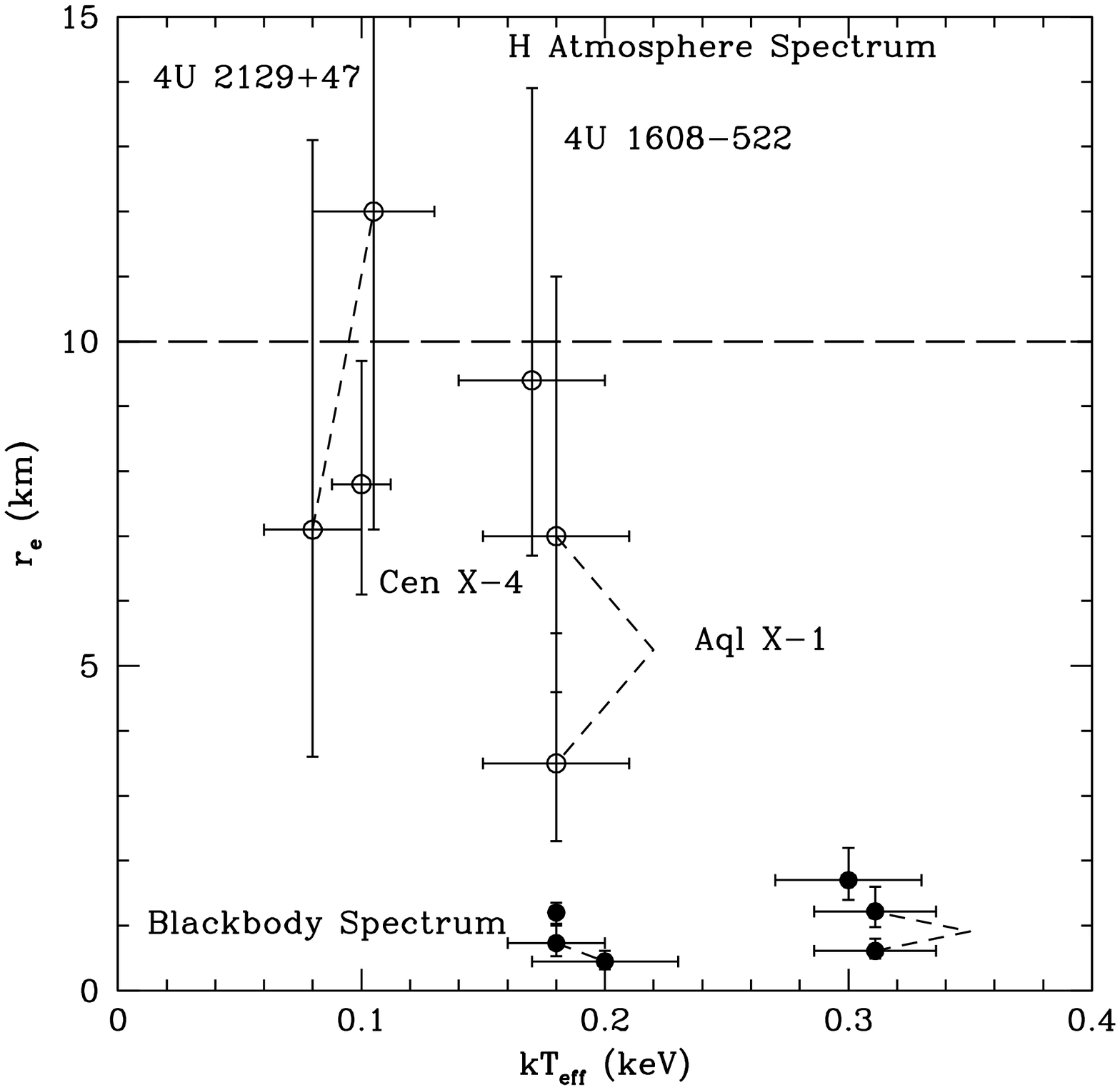 hoffset=-80 voffset=-80}{14.7cm}{21.5cm}
\FigNum{\ref{fig:nsbbhatm}}
\end{figure}

\clearpage
\pagestyle{empty}
\begin{figure}[htb]
\PSbox{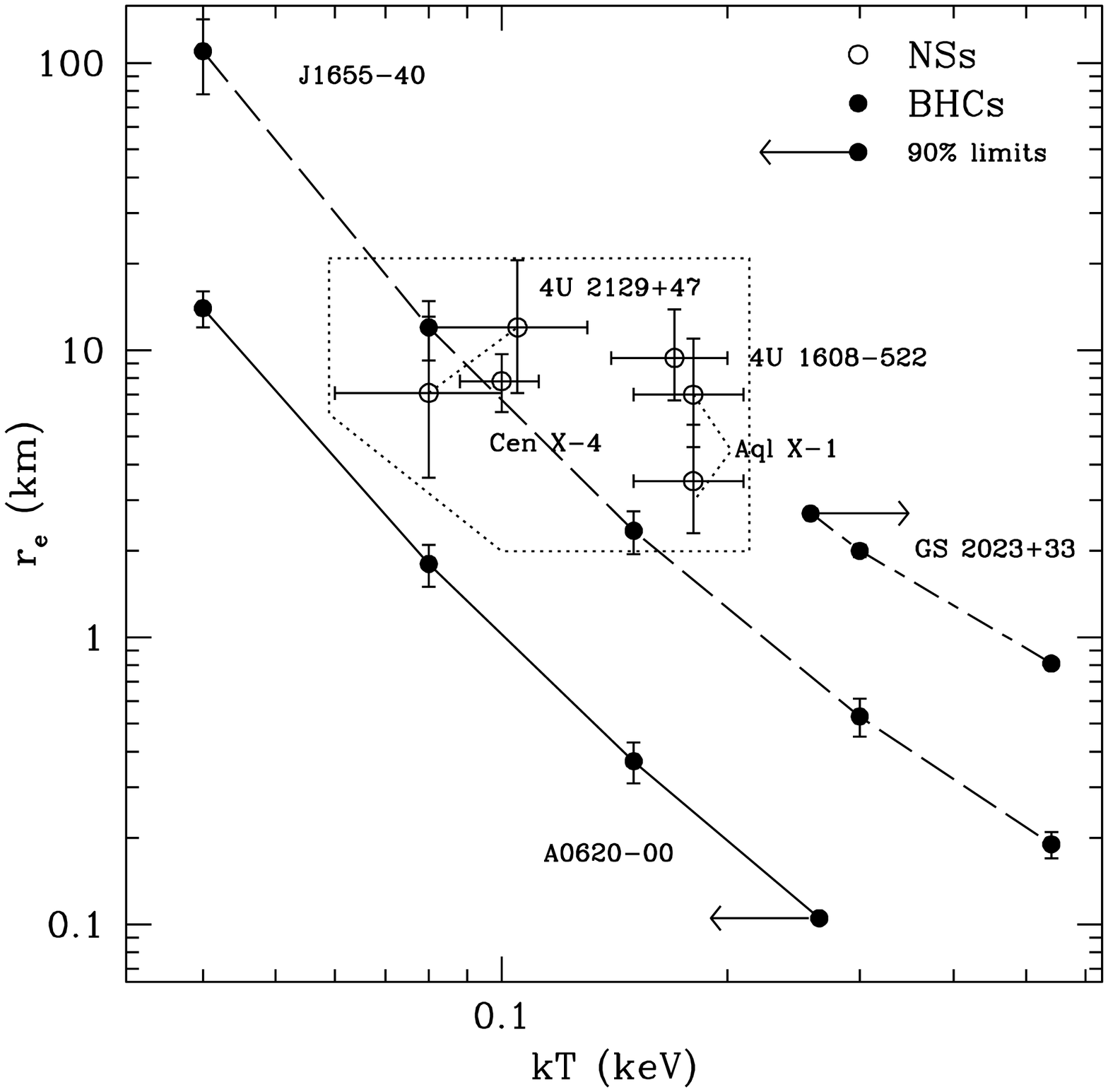 hoffset=-80 voffset=-80}{14.7cm}{21.5cm}
\FigNum{\ref{fig:nsbhcs}}
\end{figure}

\clearpage
\pagestyle{empty}
\begin{figure}[htb]
\PSbox{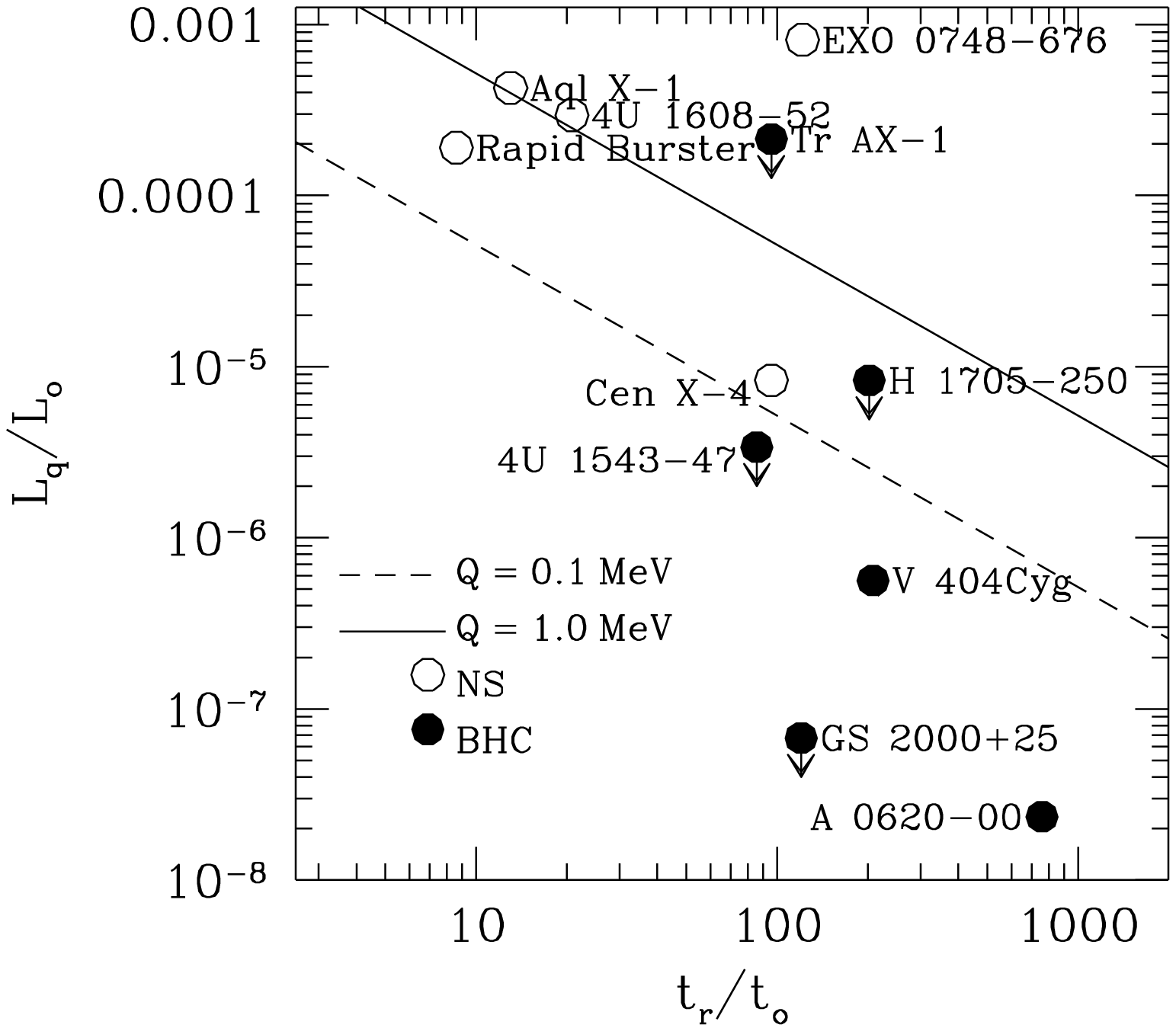 hoffset=-80 voffset=-80}{14.7cm}{21.5cm}
\FigNum{\ref{fig:Lq}}
\end{figure}

\clearpage
\newpage


\begin{table}[htb]
\begin{small}
\begin{center}
\caption{Observations \label{tab:datasets}}
\begin{tabular}{lcccc} \tableline
Number	& Satellite/Instrument&	Obs Start Time		& Live Time 	& Avg Countrate\tablenotemark{a} \\
	&		      & (UT)			& (ksec)	& (c/s) \\ \tableline  \tableline
\multicolumn{5}{c}{\nper}\\
1 	&\rosat/PSPC	& 17 Aug 1993 04:37 		& 28.0		& (8.0\ppm0.2)\tee{-2}		\\
2  	&\rosat/HRI	& 6 Jun 1995 12:29		& 18.8		& (3.3\ppm3.3)\tee{-4} \\ \tableline
\multicolumn{5}{c}{ \nmon} \\
3 	& \rosat/PSPC	& 10 Mar 1992 17:05	 	& 29.8		& (1.5\ppm0.3)\tee{-3} \\ \tableline
\multicolumn{5}{c}{ \nmus}       \\
4 	& \rosat/PSPC	& 1 Mar  1992 00:11		& 16.0		& (4.0\ppm3.5)\tee{-4} 	     \\  \tableline
\multicolumn{5}{c}{\nsco}  \\
5 	& ASCA		& 23 Mar 1996 00:54		& 87.0		& (2.1\ppm0.4)\tee{-3} (SIS1)		\\  \tableline
\multicolumn{5}{c}{ \qzvul}   \\
6 	& \rosat/PSPC	& 1 May  1992 13:23		& 12.4     	& (0.8\ppm6.0)\tee{-4}\tablenotemark{b}   \\ 
7 	& \rosat/PSPC	& 9 Apr 1993 06:41		& 6.6		& --	     \\
8 	& \rosat/PSPC	& 13 Oct 1993 13:11  		& 6.8		& --    \\  \tableline
\multicolumn{5}{c}{ \v404cyg}  \\
9 	& \rosat	& 5 Nov 1992 13:04		& 16.5		& (2.1\ppm0.1)\tee{-2} 	     \\ \tableline
\multicolumn{5}{c}{4U 2129+47}  \\
10 	& \rosat/HRI	& 9 Dec 1991 12:47		& 7.0		& (4.8\ppm1.0)\tee{-3} \\
11 	& \rosat/PSPC	& 6 Mar 1994 00:10		& 31.0		&  (6.7\ppm0.5)\tee{-3}\\ \tableline
\tablenotetext{a}{ Count-rates are background subtracted. For
\rosat/PSPC and HRI, countrates are for the energy range 0.4-2.4 keV; for \asca\ the relevant instrument is listed.}
\tablenotetext{b}{Average for observations 6, 7, \& 8}
\end{tabular}
\end{center}
\end{small}
\end{table}

\begin{table}[htb]
\begin{center}
\tablewidth{40pt}
\caption{ Object Information \label{tab:objects}}
\begin{tabular}{lccc} \tableline \tableline
Object		& $d_0$			& Mass		& \nh				\\
		& (kpc)			& (\msun)       & (\ee{22}\percm)		\\ \tableline
\multicolumn{4}{c}{Black Hole Candidates with Mass Estimates} \\
\nper		&  2.6\supp{1}		&  $>$9\supp{2}	& 0.22\supp{3}		\\ 
\nmon		&  0.9\supp{4}		&  $>$7.3\supp{5}& 0.22\supp{6,7}		\\ 
\nmus		&  3.3\supp{8}		&  5.0-7.5\supp{9} & 0.16\supp{10}\\ 
\nsco 		&  3.2\supp{11}		&  7.02\ppm0.22\supp{12}& 0.70\supp{13,14,15}		 \\ 
\qzvul		&  2.0\supp{16}		&  7.0-7.7\ppm0.5\supp{17}& 0.66-1.14\supp{13,16,18}\\ 	
\v404cyg	&  3.5 \supp{19}	&  8-12\supp{19}& 0.57\supp{20}			\\  \tableline
\multicolumn{4}{c}{Neutron Star}\\
4U 2129+47\supp{a}&  1.5\supp{21} 	& (NS)		& 0.28\supp{21}			\\
4U 2129+47\supp{b}&  6.0\supp{22} 	& (NS)		& 0.17\supp{22}			\\ \tableline
\tablerefs{
1, \citenp{esin98}; 
2,   \citenp{beekman97}; 
3, \citenp{shrader92}; 
4, \citenp{oke77}; 
5,  \citenp{mcclintock86}; 
6,  \citenp{wu76}; 
7,  \citenp{oke77}; 
8, \citenp{shahbaz97} ;
9,  \citenp{orosz96};
10,  \citenp{ebisawa94}; 
11,  \citenp{hjellmingrupen95}; 
12, \citenp{orosz97}; 
13,  W3nH,  \citenp{dickey90}; 
14,  \citenp{ueda98}; 
15,  \citenp{horne96}; 
16, \citenp{chev90}; 
17,  \citenp{casares95}; 
18, \citenp{tsunemi89}; 
19, \citenp{wagner92}; 
20,  \citenp{wagner91}; 
21,  \citenp{thorstensen79}; 
22,  \citenp{cowley90}; 
a, \citenp{thorstensen79}, \citenp{chev89b}; 
b, \citenp{cowley90}
}
\tablenotetext{}{Conversion factors: \nhtt=0.179~$A_V$ ; $A_V$=3.1~\ebv . See Text}
\end{tabular}
\end{center}
\end{table}

\begin{table}[htb]
\begin{tiny}
\begin{center}
\caption{Derived Spectral Parameters \label{tab:results}}
\begin{tabular}{lcccc} \tableline \tableline
Dataset 	&  \nh 		   &\kteff 	& $r_e$  			&$\chi^2$(dof) \\
Number		&{(\ee{22} \percm)}& {(keV)} 	& {(km ($d/d_0$))}&	\\	\tableline \tableline
\multicolumn{5}{c}{\nper}\\	
1		&  $<$0.12	   & 0.14\ppm 0.01& 3\ud{ 2.4}{0.7}	& 1.64/1 	\\
		&  (0.22)	   & 0.14\ppm0.01 &   11\ud{ 4.7}{ 2.5} & 6.3/2		\\
2 		&  (0.22)	   & (0.04)	& $<$27			& n/a		\\ 
		&  (0.22)	   & (0.08)	& $<$3.6		& n/a		\\
		&  (0.22)	   & (0.15)	& $<$0.8		& n/a		\\
		&  (0.22)	   & (0.30)	& $<$0.2		& n/a		\\
		&  (0.22)	   & (0.54)	& $<$0.08		& n/a		\\ \tableline
\multicolumn{5}{c}{{\nmon}} \\
3 		& (0.22)	   & (0.04)	& 14\ppm 2		& 1.2/2		\\
		& (0.22)	   & (0.08)	& 1.8 \ppm 0.3		& 0.3/2	\\
		& (0.22)	   & (0.15)	& 0.37 \ppm 0.06	&1.9/2		\\
		& (0.22) 	   & (0.265) 	& 0.105			& 4.0/2 \\ \tableline
\multicolumn{5}{c}{ {\nmus} } \\
4 		& (0.16)	   & (0.04)	& $<$82			&n/a		\\
		& (0.16)	   & (0.08)	& $<$8.6		&n/a		\\
		& (0.16)	   & (0.15)	& $<$1.2		&n/a		\\
		& (0.16)	   & (0.30)	& $<$0.26		&n/a		\\
		& (0.16)	   & (0.54)	& $<$0.13		&n/a		\\	\tableline
\multicolumn{5}{c}{\nsco}\\		
5		& (0.74)	   & (0.04)	& 110\ppm32		& 105/70	\\
		& (0.74)	   & (0.08)	& 12.0\ppm2.8		& 100/70	\\
		& (0.74)	   & (0.15)	& 2.35\ppm0.4		& 95/70	\\
		& (0.74)	   & (0.30)	& 0.53\ppm0.08		& 80/70	\\
		& (0.74)	   & (0.54)	& 0.19\ppm0.02		& 61/70	\\ \tableline
\multicolumn{5}{c}{\qzvul} \\
6, 7, \& 8	& (0.66)	   & (0.04)	& $<$28			&n/a		\\
		& (0.66)	   & (0.08)	& $<$3.8		&n/a		\\
		& (0.66)	   & (0.15)	& $<$1.0		&n/a		\\
		& (1.1)		   & (0.15)	& $<$1.9		&n/a		\\
		& (0.66)	   & (0.30)	& $<$0.29		&n/a		\\
		& (0.66)	   & (0.54)	& $<$0.13		&n/a		\\	\tableline
\multicolumn{5}{c}{\v404cyg}\\		
9 		& (0.54)	   & (0.258)	& 2.7\ppm0.1		& 4.0/2		\\	
		& (0.54)	   & (0.30)	& 2.0\ppm0.1		& 1.8/2		\\	
		& (0.54)	   & (0.54)	& 0.81\ppm0.03		& 0.72/2		\\	\tableline
\multicolumn{5}{c}{4U 2129+47 (1.5 kpc; \nhtt=0.28)}\\		
11		& (0.28)	   & 0.08\ud{0.02}{0.015}&7.1\ud{6.0}{3.5}& 0.13/1	\\
\multicolumn{5}{c}{4U 2129+47 (6.0 kpc; \nhtt=0.17)}\\		
11		& (0.17)	   & 0.105\ud{0.025}{0.020}&12\ud{8.6}{4.9}& 0.71/1	\\
\tablenotetext{}{ Errors and upper-limits are 90\% confidence; values listed in
parenthesis were fixed during the fits. For description of the different ($d_0$, \nh) assumptions for 4U~2129+47, see App.~\protect\ref{sec:2129} }
\end{tabular}
\end{center}
\end{tiny}
\end{table}

\begin{table}[htb]
\begin{center}
\caption{Bolometric Luminosity of Transient Neutron Stars in Quiescence\label{tab:lbol}}
\begin{tabular}{lcccc} \tableline \tableline
Source			&	$L_x$ (0.5-10.0 keV)			& $L_{\rm bol}$ \\
			&	\ee{32} \cgslum\			& \ee{32} \cgslum\ 	\\ \tableline
Aql X-1			&  (20.4\ppm9.2) $(d_0/4 \kpc)^{2}$ 	&	39	\\
Cen X-4			&  (1.6\ppm0.6) $(d_0/1.2 \kpc)^2$ 	&	4.6	\\
\x1608\ 		&  (30.0\ppm15.2) $(d_0/3.6 \kpc)^2$	&	56	\\
4U 2129+47\supp{a}	&  (0.7\ud{1.5}{0.3}) $(d_0/1.5 \kpc)^2$	(0.5-3.0 keV)& 1.6	\\
4U 2129+47\supp{b}	&  (5.6\ud{12}{3})  $(d_0/6.0 \kpc)^2$ (0.5-3.0 keV)& 13	\\ \tableline
\tablenotetext{}{\supp{a,b} A range of (\nh , $d_0$) is assumed; See App.~\protect\ref{sec:2129}}
\end{tabular}
\end{center}
\end{table}

\end{document}